\documentclass[aps,twocolumn, superscriptaddress]{revtex4-1}
\usepackage[pdftex]{graphicx}% Include figure files
 \usepackage{amsmath}
\usepackage[T1]{fontenc}
\usepackage{amsfonts}
\usepackage{lipsum}
\usepackage{amssymb, mathrsfs}
\usepackage{braket}
\usepackage{subfigure}
\def\beq{\begin{equation}}
\def\eeq{\end{equation}}
\def\bsp{\begin{split}}
\def\esp{\end{split}}
\def\bea{\begin{eqnarray}}
\def\eea{\end{eqnarray}}
\def\ba{\begin{array}}
\def\ea{\end{array}}

\def\lb{\left(}
\def\rb{\right)}

\def\l.{\left.}
\def\r.{\right.}

\def\ra{\rangle}
\def\la{\langle}

\begin{document}

\title{Topological Thermal Hall Effect  in  Frustrated Kagom\'e Antiferromagnets}
\author{S. A. Owerre}
\affiliation{Perimeter Institute for Theoretical Physics, 31 Caroline St. N., Waterloo, Ontario N2L 2Y5, Canada.}
\affiliation{African Institute for Mathematical Sciences, 6 Melrose Road, Muizenberg, Cape Town 7945, South Africa.}

\begin{abstract}
 In frustrated magnets the Dzyaloshinsky-Moriya  interaction (DMI) arising from spin-orbit coupling (SOC) can induce a magnetic long-range  order.  Here, we report a theoretical prediction of thermal Hall effect in frustrated kagom\'e magnets such as  KCr$_3$(OH)$_6$(SO4)$_2$  and KFe$_3$(OH)$_{6}$(SO$_{4}$)$_2$. The thermal Hall effects in these materials are induced by scalar spin chirality as opposed to DMI in previous studies.  The scalar spin chirality originates from  magnetic-field-induced chiral spin configuration due to non-coplanar spin textures, but in general it can be spontaneously developed as a macroscopic order parameter in chiral quantum spin liquids. Therefore, we infer that there is a possibility of  thermal Hall effect  in frustrated kagom\'e magnets such as herbertsmithite ZnCu$_3$(OH)$_6$Cl$_2$ and the chromium compound Ca$_{10}$Cr$_7$O$_{28}$, although they also  show evidence of magnetic long-range order in the presence of applied magnetic field or pressure.   
\end{abstract}
\maketitle

\textit{Introduction}--. Topological phases of matter are an active research field in condensed matter physics, mostly dominated by electronic systems. Quite recently  the concepts of topological matter have been extended to nonelectronic bosonic systems such as  quantized spin waves (magnons) \cite{shin1, alex1, alex1a, alex0, alex2,alex5,alex4, sol1,sol, alex5a, alex6,  sol4,  mok,su,fyl,lifa,bol} and quantized lattice vibrations (phonons) \cite{pho,pho1,pho2,pho3,pho4,pho5}.   In the former, spin-orbit coupling  manifests in the form of Dzyaloshinsky-Moriya  interaction \cite{dm,dm2} and it leads to  topological spin excitations and chiral edge modes in collinear ferromagnets \cite{lifa,alex4}. The quantized spin waves or magnons are charge-neutral quasiparticles and they do not experience a Lorentz force as in charge particles. However,  a temperature gradient can induce a heat current and the DMI-induced Berry curvature acts as an effective magnetic field in momentum space. This leads to a thermal version of the Hall effect characterized by a temperature dependent thermal Hall conductivity  \cite{alex0,alex2}.     Thermal Hall effect  is now an emerging active research area for probing the topological nature of magnetic spin excitations in quantum magnets.  The thermal Hall effect of spin waves has been realized experimentally in a number of pyrochlore ferromagnets \cite{alex1,alex1a}. Recently,  DMI-induced topological magnon bands and thermal Hall effect have been observed  in collinear kagom\'e ferromagnet Cu(1-3, bdc) \cite{alex5a,alex6}. 
 
In frustrated kagom\'e magnets, however, there is no magnetic long-range order (LRO) down to the lowest accessible temperatures. The classical ground states have an extensive degeneracy and they are considered as  candidates for quantum spin liquid (QSL) \cite{Sav,nor}. The ground state of spin-$1/2$ Heisenberg model on the kagom\'e lattice is believed to be a U(1)-Dirac spin liquid \cite{ran}. In physical realistic materials, however, there are other interactions and perturbations that tend to alleviate QSL ground states and lead  to LRO.   Recent experimental syntheses  of kagom\'e antiferromagnetic materials have shown that the effects of SOC or DMI are not negligible in frustrated magnets. The DMI is an intrinsic perturbation to the Heisenberg interaction which arises from SOC and it affects the low-temperature physics of frustrated magnets. One of the striking  features of the DM perturbation  is that it can induce LRO with a $\bold q=0$ propagation vector in frustrated  kagom\'e magnets \cite{men1}.  Hence, the DMI suppresses the QSL phase of frustrated kagom\'e antiferromagnets up to a critical value \cite{men3}. The syntheses of materials have shown that  various experimentally accessible  frustrated  kagom\'e antiferromagnets show evidence of coplanar/noncollinear $\bold q=0$ LRO  at specific temperatures  \cite{sup1a, men1, men3, men4a,gro2,gro4}. The famous frustrated kagom\'e magnets with this LRO are jarosites, such as KCr$_3$(OH)$_6$(SO4)$_2$ \cite{gro2,gro4}  and KFe$_3$(OH)$_{6}$(SO$_{4}$)$_2$ \cite{men1,sup1a}.  Even  highly frustrated magnets with QSL ground states such as herbertsmithite ZnCu$_3$(OH)$_6$Cl$_2$ \cite{zor} and Ca$_{10}$Cr$_7$O$_{28}$ \cite{balz} are fragile in the presence of applied  magnetic field or pressure and they show evidence of LRO \cite{jeo,koz}.  However, the role of DMI and magnetic field in frustrated kagome magnets has not been investigated in the context of thermal Hall effect and topological spin excitations. 

 Usually, the thermal Hall effect of spin excitations arises in insulating ferromagnets because the spontaneous magnetization combined with DMI break the time-reversal symmetry (TRS) macroscopically even in the absence of applied magnetic field. In this report, we show that the DMI is not the primary source of thermal Hall effect  in frustrated kagom\'e magnets  with/without LRO. Rather, a field-induced scalar spin chirality arising from the non-coplanar chiral spin configuration yields both thermal Hall effect and topological spin excitations. We note that in general the presence of  scalar spin chirality does not necessarily require a LRO or applied magnetic field as in chiral QSLs \cite{kal,wen,bas,bau}. Therefore our findings can be extended to a wide range of disordered QSL phases on the kagom\'e lattice in which TRS is broken spontaneously and macroscopically.   The experimental probe for the thermal Hall effect in these frustrated magnets will provide an understanding of both topological magnetic excitations and scalar spin  chirality. These materials can be useful for designing systems with low-dissipation  applicable to spin-based computing or magnon spintronics \cite{magn}.  In this respect, our results sharply contrast with collinear magnetic systems \cite{alex6, alex1,alex5a} and triplon excitations in dimerized quantum magnets \cite{jud} in the presence of applied magnetic field with finite DMI, but zero scalar spin  chirality.  \\

\begin{figure}
\includegraphics[width=3.2in]{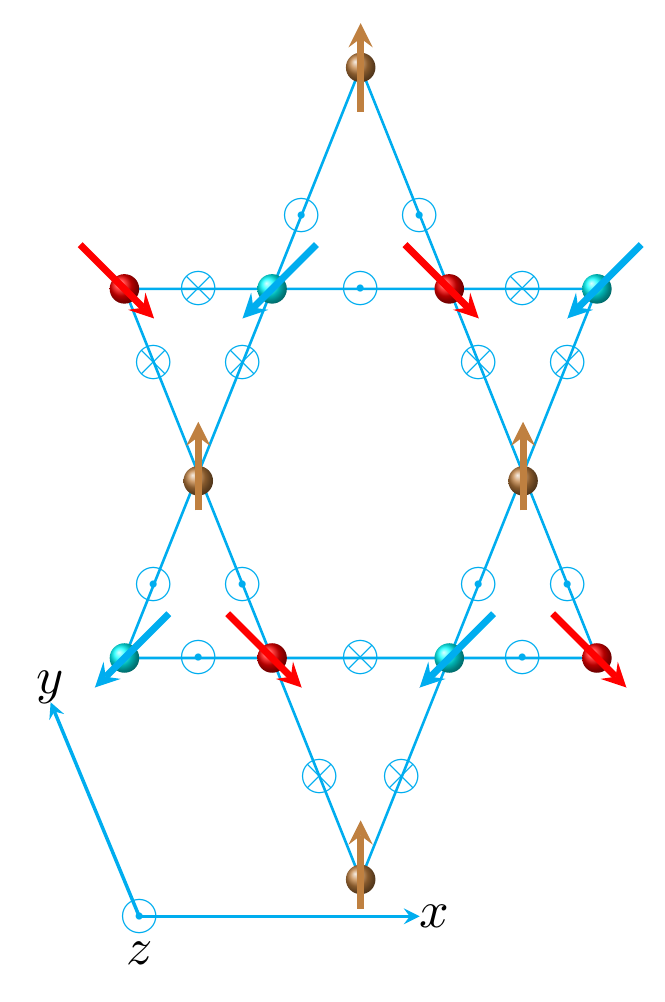}
\caption{{\bf Kagom\'e lattice}.  Coplanar $\bold q=0$ spin configuration on the kagom\'e lattice with positive vector chirality. The alternating  DMI lies at the midpoint between two magnetic ions.}
\label{app1}
\end{figure}

\textit{Model}--. 
We consider the Hamiltonian for frustrated kagom\'e antiferromagnets given by \begin{align}
\mathcal H&= \mathcal J\sum_{\la i,j\ra }{\bf S}_{i}\cdot{\bf S}_{j}+\mathcal J_2\sum_{\la\la i,j\ra\ra}{\bf S}_{i}\cdot{\bf S}_{j} +\sum_{\la i,j\ra } \boldsymbol{\mathcal{D}}_{ij}\cdot{\bf S}_{i}\times{\bf S}_{j}.
\label{h}
\end{align}
The first summation runs over nearest-neighbours (NN) and the second runs over next-nearest-neighbours (NNN), where $\mathcal J,\mathcal J_2>0 $ are the isotropic antiferromagnetic couplings respectively and  ${\bf S}_{i}$ is the spin moment at site $i$.  According to Moriya rules \cite{dm2}, the midpoint between two magnetic ions on the kagom\'e lattice is not an inversion center. Therefore  the DM vector $\boldsymbol{\mathcal{D}}_{ij}$  can be allowed on the kagom\'e lattice as shown in Fig.~\ref{app1}.  In ferromagnets the DMI breaks TRS and leads to topological spin waves \cite{shin1, alex1, alex1a, alex0, alex2,alex5,alex4,bol, sol1,sol, alex5a, alex6,  sol4,  mok,su,fyl,lifa}, whereas in the present model the DMI  is known to stabilize the 120$^\circ$ coplanar  $\bold q=0$ magnetic structure \cite{men1}. Besides,  the NNN coupling $\mathcal J_2>0$  can equally  stabilizes the coplanar   magnetic structure \cite{har1}. The out-of-plane  DMI $\boldsymbol{\mathcal D}_{ij}=(0,0, \mp \mathcal D_z)$ is intrinsic to the kagom\'e lattice, where $-/+$ alternates between up/down triangles of the kagom\'e lattice as shown in Fig.~\ref{app1}. The sign of out-of-plane DMI determines the vector chirality of the coplanar 120$^\circ$ order and  only one ground state is selected for each sign of the DMI \cite{men1}. The positive vector chirality  in Fig.~\ref{app1}  with $\mathcal D_z>0$ is the ground state of most jarosites and we will consider this case. The DMI breaks the global SO(3) rotation symmetry of the Hamiltonian down to SO(2) global spin rotation symmetry in the $x$-$y$ plane.  Depending on the frustrated kagom\'e magnet the in-plane DM component may vanish or negligible \cite{gro2,gro4,zor, men3}. When it is present, it breaks mirror reflection symmetry of the lattice and global spin rotation symmetry. It can  lead to spin canting with weak out-of-plane ferromagnetic moment. However,  most materials have  very small or negligible in-plane DM components  due to dominant out-of-plane components \cite{gro2,gro4,zor, men3}. Therefore its presence will not change the basic results of this report.
\begin{figure*}
\centering
\includegraphics[width=3.2in]{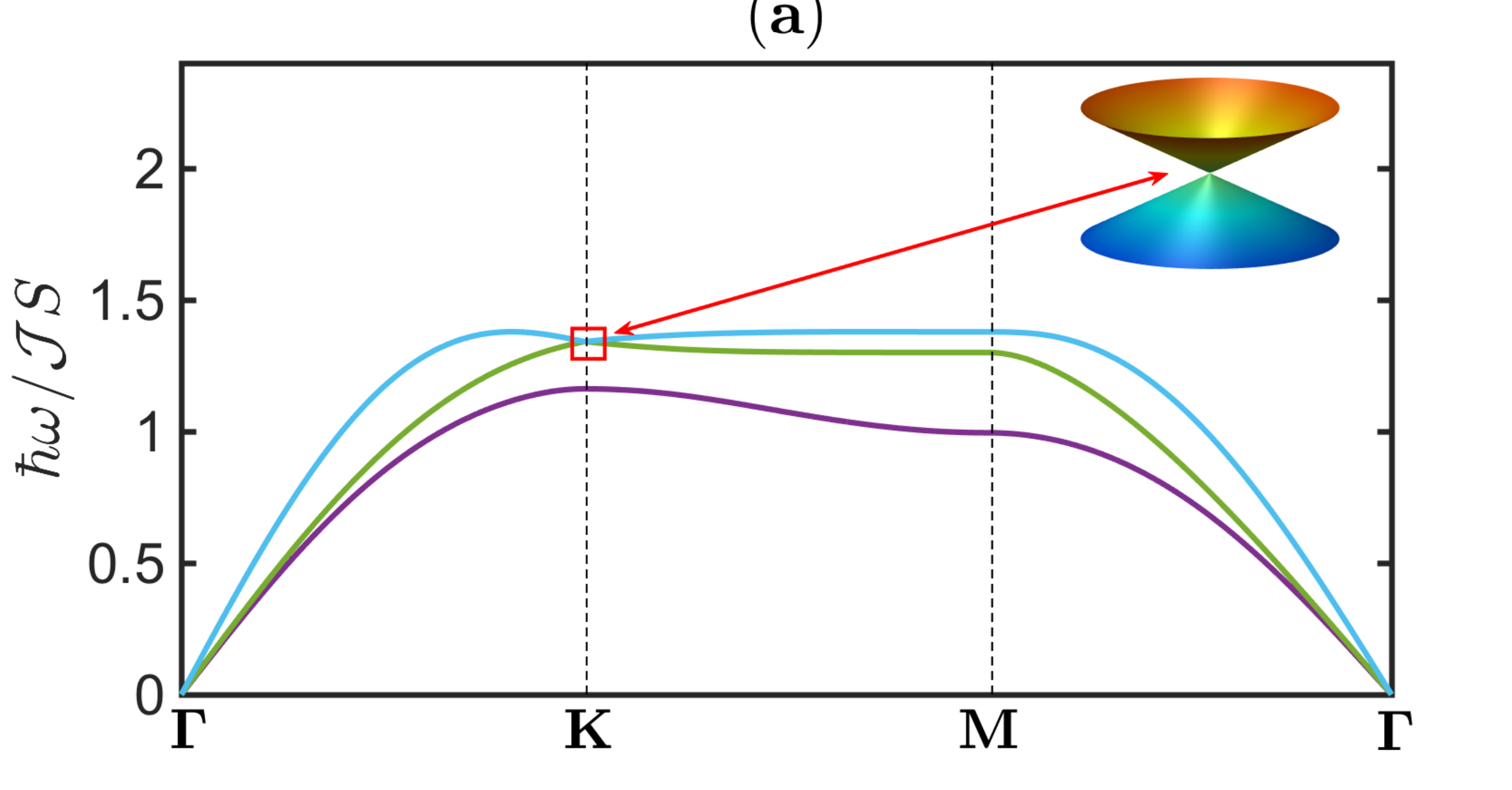}
\includegraphics[width=3.2in]{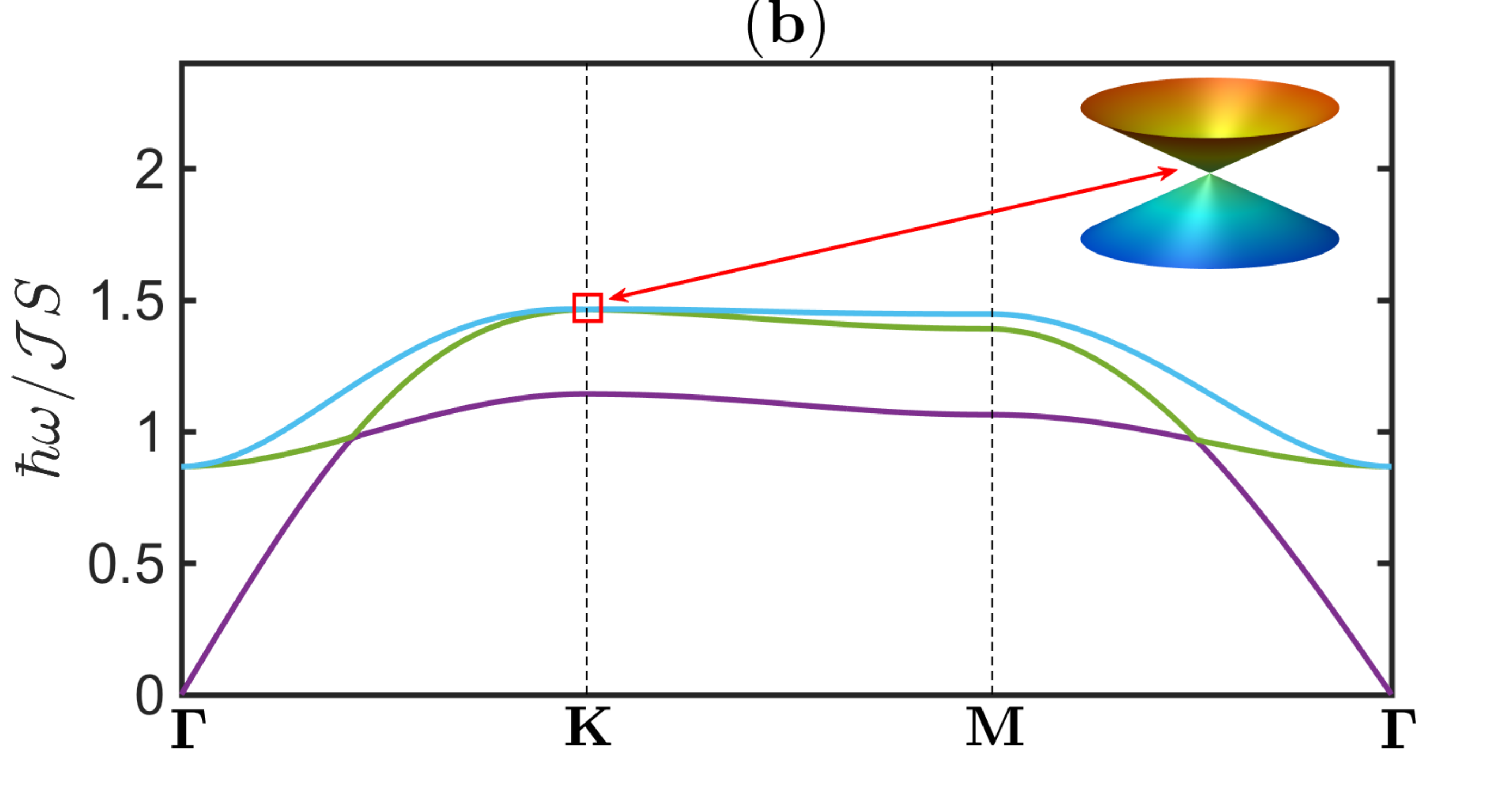}
\caption{{\bf Spin wave excitations at zero magnetic field}.  ${\bf a},~\mathcal{D}/\mathcal{J}=0, \mathcal{J}_2/\mathcal{J}=0.3$. ${\bf b},~\mathcal{D}/\mathcal{J}=0.2, \mathcal{J}_2/\mathcal{J}=0.1$. The inset shows gapless Dirac cone at ${\bf K}$. }
\label{dm}
\end{figure*}
\begin{figure*}
\centering
\includegraphics[width=3.2in]{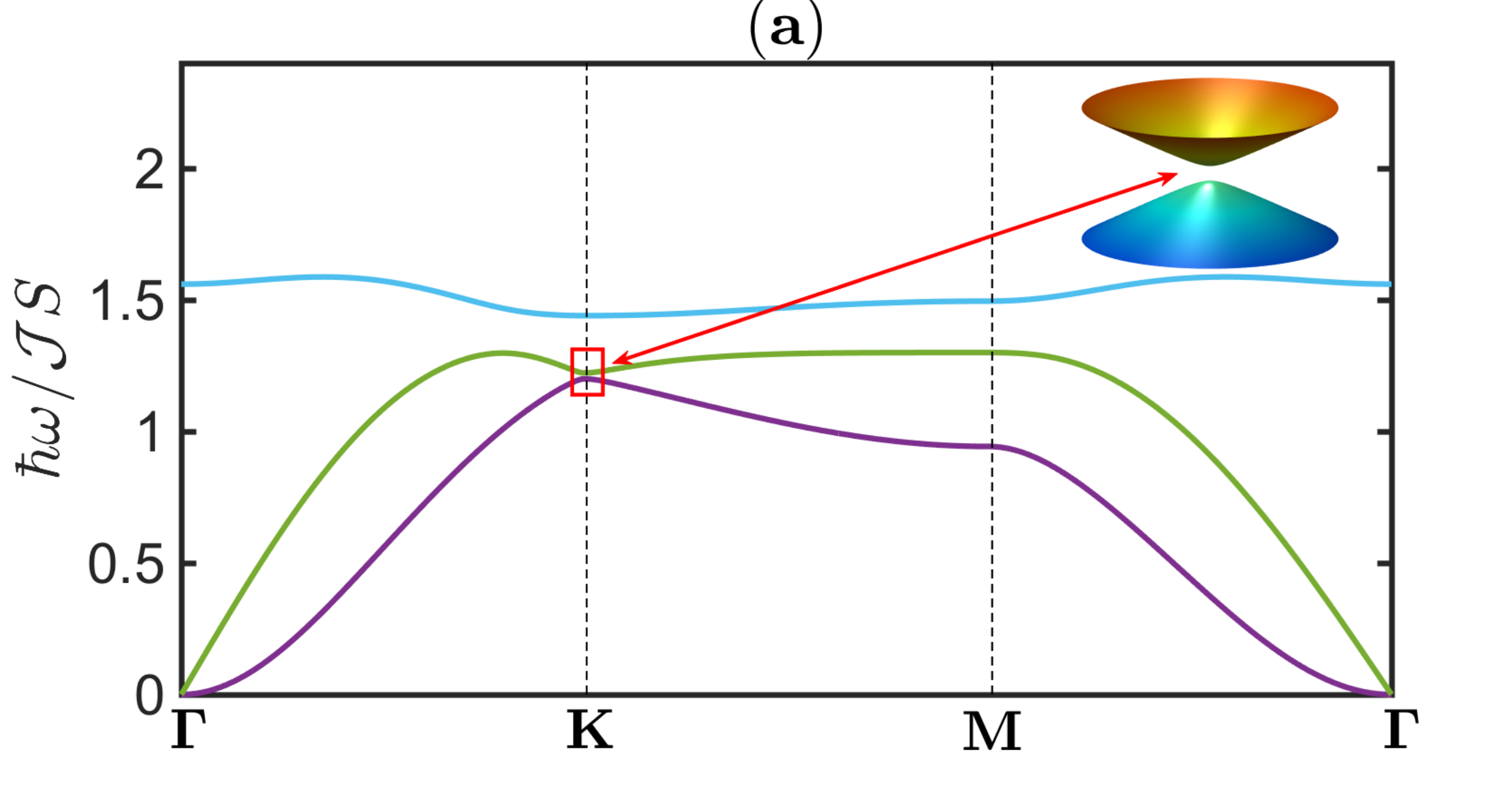}
\includegraphics[width=3.2in]{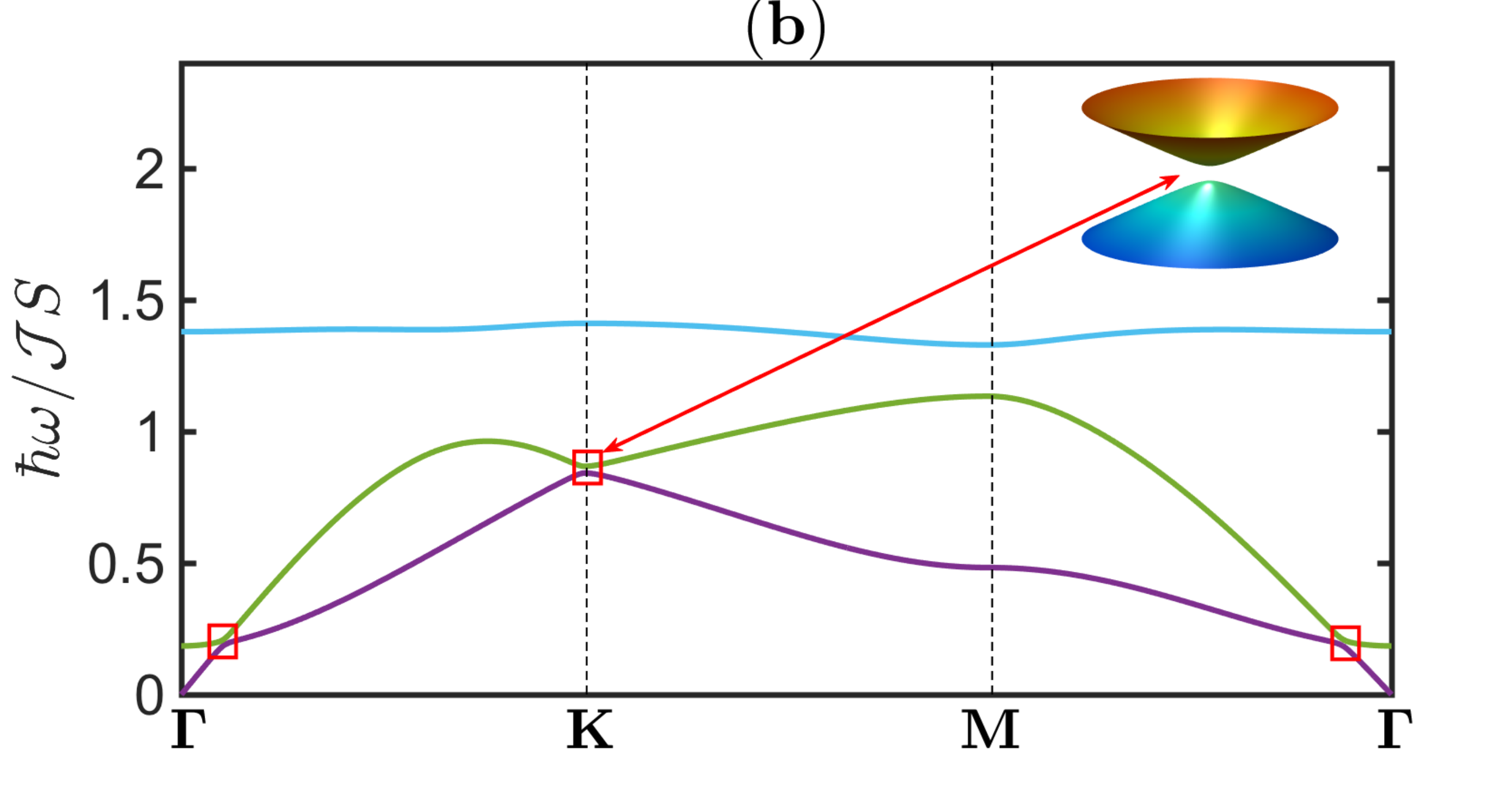}
\caption{{\bf Topological spin wave excitations at nonzero magnetic field} $h=0.4 h_s$.  ${\bf a}$, $\mathcal D_z/\mathcal J=0,~\mathcal J_2/\mathcal J=0.3$.  ${\bf b}$, $\mathcal D_z/\mathcal J=0.06,~\mathcal J_2/\mathcal J=0.03$ with  $\mathcal J=3.34 ~\text{meV}$. The inset corresponds to gap magnon bands indicated by red squares.}
\label{band3}
\end{figure*}

\textit{Dirac magnon}--. As we mentioned above, the presence of sizeable DMI in this model has been proven to induce LRO. Nevertheless, the present model still differs significantly from collinear magnets  because the DMI  does not play the same role in both systems as we will show. Figure \ref{dm} shows the spin wave excitations of the 120$^\circ$ coplanar spins (see Ref.~\cite{SM}) along the high symmetry points of the Brillouin zone \cite{owee} with ${\bf \Gamma}=(0,0)$, ${\bf M}=(\pi/2,\pi/2\sqrt{3})$ and ${\bf K}=(2\pi/3,0)$.  For $\mathcal J_2\neq 0,\mathcal D_z= 0$ the global SO(3) rotation symmetry is restored  and the three dispersive bands have Goldstone modes at the ${\bf \Gamma}$ point. The zero energy mode usually present in the kagom\'e antiferromagnet is lifted by $\mathcal J_2>0$.  For $\mathcal J_2\neq 0,\mathcal D_z\neq 0$  the SO(3) rotation symmetry is broken down to SO(2) in the $x$-$y$ plane giving rise to one Goldstone mode at the ${\bf \Gamma}$ point. The  zero energy mode is also lifted to a constant energy mode by the DMI, which acquires a small dispersion due to $\mathcal J_2>0$.    The interesting feature of this model is the  linear band crossing of two magnon branches at the ${\bf K}$ point, which realizes a two-dimensional (2D) Dirac Hamiltonian
\begin{align}
\mathcal H(\pm{\bf K}+ {\bf q})= c_0\mathbb{I}_{2\times 2} + c_1\lb \pm q_x\sigma_x +q_y\sigma_y\rb,
\label{dirac}
\end{align}
where $\bold q$ is the momentum vector, $c_0$ and $c_1$ are functions of the Hamiltonian parameters, $\boldsymbol{\sigma}$ is a Pauli matrix and $\mathbb{I}_{2\times 2}$ is an  identity $2\times 2$ matrix.  The linearized Hamiltonian \eqref{dirac}  has winding number $\pm 1$ for a closed loop encircling the states at $\pm{\bf K}$. This linear band crossing is different from the Goldstone modes of continuous rotational symmetry breaking.   The persistence of Dirac points in the presence of SOC (in this case DMI) is the basis of Weyl magnon \cite{mok,su} and Dirac semimetal in electronic systems \cite{ste}. In this regard, the present model can be deemed a magnon analog of quasi-2D Dirac semimetal.    From the symmetry point of view, the mirror reflection symmetry with respect to the kagom\'e planes is a good symmetry of the kagom\'e lattice, but reverses the 120$^\circ$ coplanar spins and TRS brings the spins back to the original states.  Hence, the combination of  mirror reflection and time-reversal symmetries leaves  the 120$^\circ$ coplanar spins invariant.  Therefore, the Dirac points are protected by this combined symmetry. A thorough study of Dirac magnon points with DMI is beyond the purview of this report and will be presented in detail elsewhere. For Fe-jarosites a small in-plane DMI induces a gap at the ${\bf \Gamma}$ point  and very small gap at the ${\bf K}$ point,  but it does not remove the  linear band crossing at ${\bf \Gamma}-{\bf K}$ line \cite{men4a}. The point here is that the DMI does not lead to topological magnon bands unlike in the previous studies.

\begin{figure}
\centering
\includegraphics[width=3.5in]{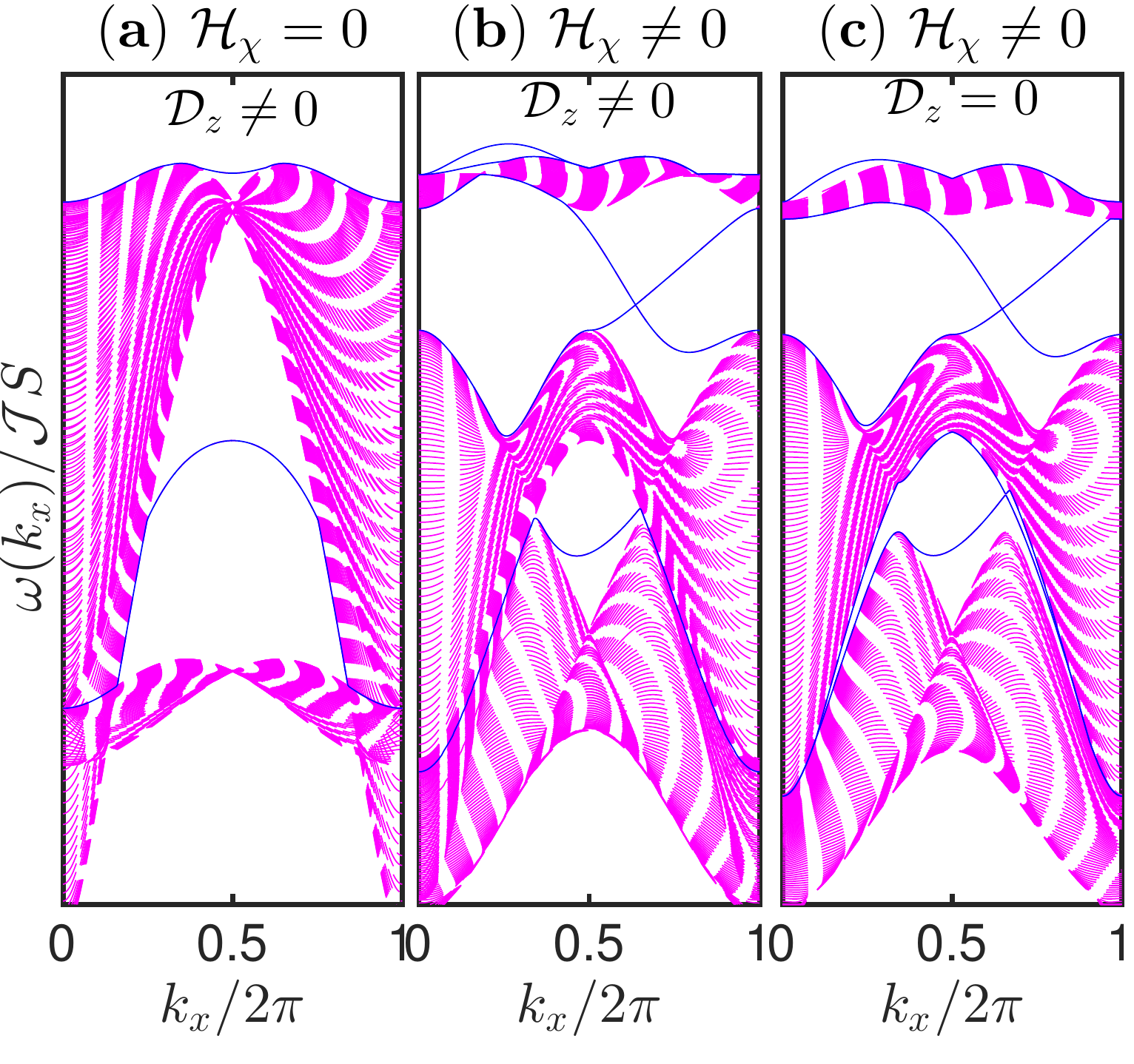}
\caption{{\bf Chiral spin wave edge modes}. {\bf a}, $h/h_s=0~(\mathcal H_\chi=0)$ with  DMI.  {\bf b}, $h/h_s=0.4~(\mathcal H_\chi\neq 0)$ with  DMI. {\bf c}, $h/h_s=0.4~(\mathcal H_\chi\neq 0)$  without DMI. The parameters are the same as Fig.~\ref{band3}. The pink lines are the bulk bands and the blue lines are the edge modes.}
\label{edge}
\end{figure}
\begin{figure*}
\centering
\includegraphics[width=3in]{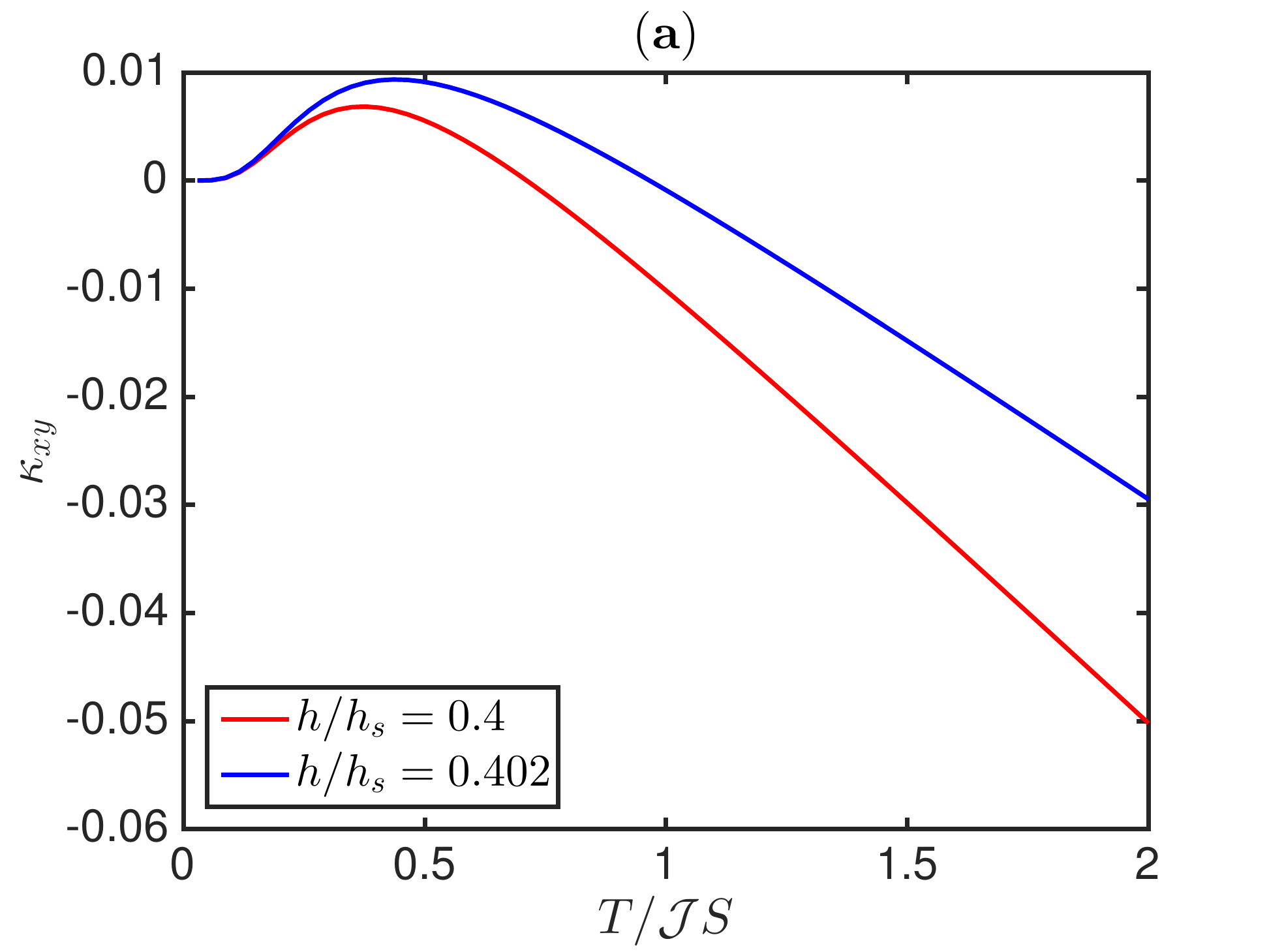}
\includegraphics[width=3in]{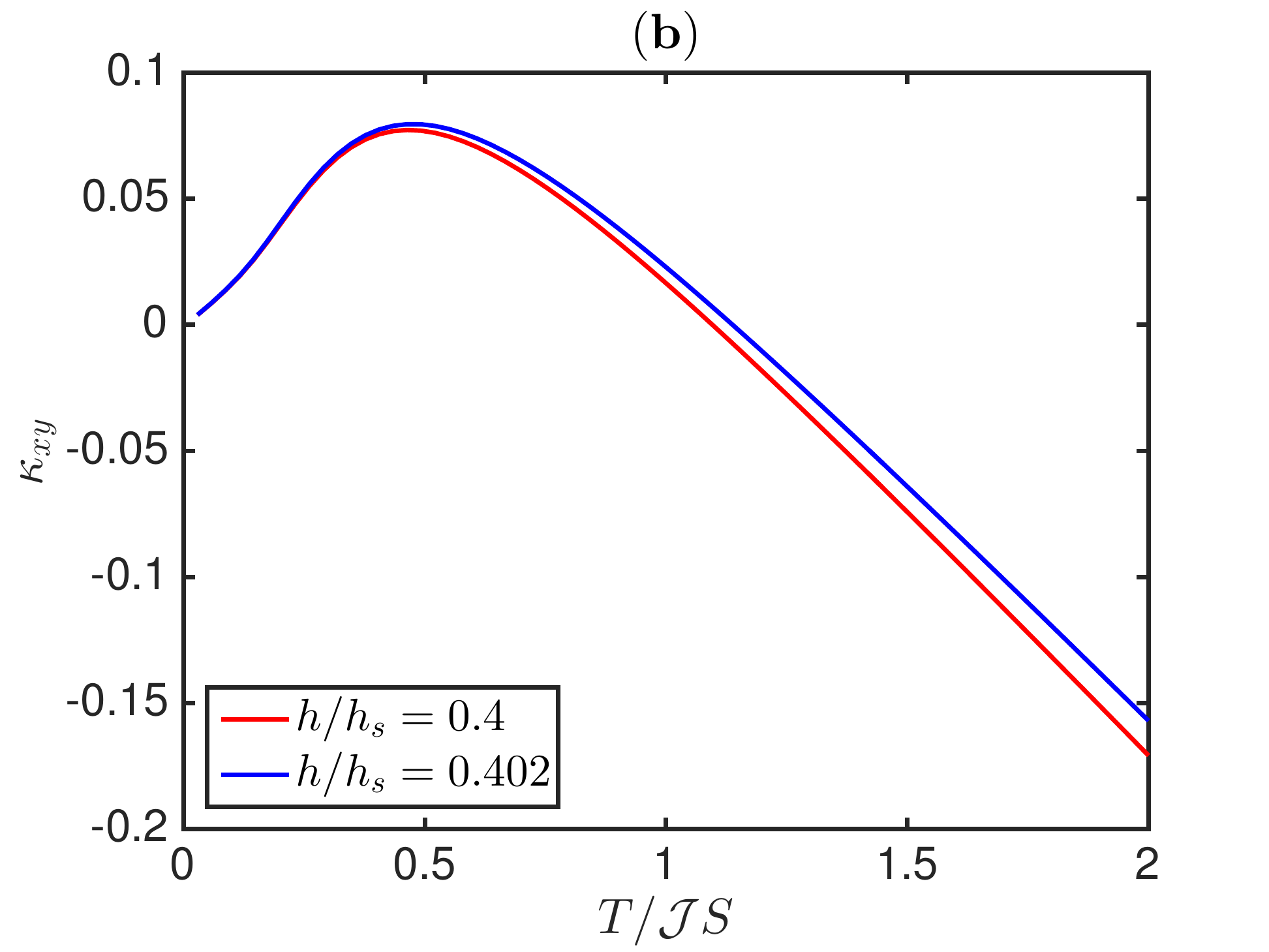}
\caption{{\bf Topological thermal Hall effect}. Low-temperature dependence of thermal Hall conductivity $\kappa_{xy}$. ${\bf a}$, $\mathcal D_z/\mathcal J=0.06,~\mathcal J_2/\mathcal J=0.03$. ${\bf b}$, $\mathcal D_z/\mathcal J=0,~\mathcal J_2/\mathcal J=0.3$.}
\label{N_th}
\end{figure*}
\textit{Topological magnon}--.  We expect the Dirac magnon points  to be at the boundary between  topological and trivial insulators just like in electronic systems. Therefore, the system can be driven to a topological phase by breaking the symmetries that protect the Dirac magnon points. As we previously mentioned, an in-plane DMI breaks mirror reflection symmetry but preserves TRS. If the in-plane DM component is the dominant anisotropy then it is capable of inducing  topological magnon bands. However, due to a dominant out-of-plane DM component in most kagom\'e antiferromagnetic crystals, the effects of the in-plane DM component is usually suppressed  \cite{men3,zor,gro2,gro4} and can be  neglected. This means that the Dirac magnon points will persist.  The system can only be driven to a topological phase by an external factor such as an external  magnetic field applied perpendicular to the kagom\'e plane, which can be written  as $
\mathcal H_{Z}=-\boldsymbol{\mathcal{B}}\cdot\sum_{i} {\bf S}_{i}$,
where $\boldsymbol{\mathcal{B}}=g\mu_B\mathcal B_z\bold{\hat z}$,  $g$ is the $g$-factor and $\mu_B$ is the Bohr magneton.  
The out-of-plane Zeeman magnetic field has a profound effect on frustrated magnets with QSL phases as it can induce a LRO \cite{jeo,koz, balz}. In the ordered regime we find that the Zeeman magnetic field induces a noncoplanar magnetic spin texture with an emergent scalar spin chirality given by (see Ref.~\cite{SM})  
\bea 
\mathcal H_{\chi}\sim\cos\chi \sum {\bold S}_i\cdot\lb \bold S_{j}\times\bold S_{k}\rb,\eea 
where $\cos\chi=h/h_s$ with $h_s= [6(\mathcal J+\mathcal J_2)+2\sqrt{3}\mathcal D_z]$ and $h=g\mu_B\mathcal B_z$.  Under $180^\circ$ rotation of the spins, the scalar chirality changes sign that is  $\mathcal H_{\chi}\to-\mathcal H_{\chi}$ as $\chi\to \pi+\chi$.

 The most important feature of this model is that the scalar spin chirality does not necessarily require the presence of DMI.  This suggests that the DMI is not the primary source of topological spin excitations in frustrated magnets, which sharply differs from collinear magnets \cite{alex1, alex1a, alex0, alex2,alex5,alex4, bol, sol1,sol, alex5a, alex6,  sol4,  mok,su,fyl,lifa, shin1} and triplon excitations \cite{jud} in a magnetic field with zero scalar spin  chirality. We note that scalar spin chirality plays a crucial role in different areas of physical interest. Most importantly, it is the  hallmark of  chiral QSLs in which TRS is broken macroscopically without any LRO, that is $\la {\bf S}_j\ra=0$ \cite{kal,wen,bas,bau}. The \textit{cuboc1} phase on the frustrated kagom\'e lattice also has a finite scalar spin chirality even in the absence of an explicit DMI \cite{gong}. The main origin of the scalar spin chirality is geometrical spin frustration. In the momentum space \cite{hp}, it is defined as a fictitious magnetic flux $\Phi$ acquired by the propagation of magnon around a set of three non-coplanar magnetic moments. This  gives rise  to  a Berry curvature  (see Ref.~\cite{SM}) which is related to the solid angle subtended by three non-coplanar spins on each triangular plaquette of the kagom\'e lattice. 
 
 The magnon energy branches of the noncoplanar spin textures are shown in Fig.~\ref{band3}  with the parameter values of KFe$_3$(OH)$_{6}$(SO$_{4}$)$_2$ \cite{men4a}.  The magnon  dispersions show a finite gap at all points in the Brillouin zone with/without DMI. The linearized Hamiltonian for two avoided band crossings  at $\pm{\bf K}$  can be written as 
  \begin{align}
\mathcal H(\pm{\bf K}+ {\bf q})= c_0\mathbb{I}_{2\times 2} + c_1\lb \pm q_x\sigma_x +q_y\sigma_y\rb + m(\Phi)\sigma_z,
\label{dirac1}
\end{align}
 where $m(\Phi)\propto\Phi$ and  $\sin\Phi$ is related to the scalar spin chirality.
In principle, a gap Dirac point is not enough to prove that a system is topological. To further substantiate the topological nature of the system we  have solved  for the chiral edge modes using a strip geometry with open boundary conditions along the $y$ direction  and infinite along $x$ direction  as depicted in Fig.~\ref{edge}. At zero magnetic field $\mathcal H_{\chi}=0$, there are no gapless edge modes, but a single edge mode connects the Dirac magnon points. As the magnetic field is turned on  $\mathcal H_{\chi}\neq 0$, we clearly see gapless edge modes between  the upper and middle bands, which signify a strong topological magnon insulator \cite{lifa}. Because of the bosonic nature of magnons there is no Fermi energy or completely filled bands in this system. Nevertheless, a Chern number can still be defined (see Ref.~\cite{SM}). The Chern numbers in the topological regime are calculated as $[0,-\text{sgn}(\sin\Phi),\text{sgn}(\sin\Phi)]$ for the lower, middle, and upper bands respectively.  This confirms that the system is in the topological phase. 
 
\textit{Topological thermal Hall effect}--. Having established the topological nature of the system,  now we will investigate an experimentally accessible aspect of insulating frustrated kagom\'e quantum magnets.  The existence of nontrivial topological spin excitations can be  probed by inelastic neutron scattering experiment via the measurement of thermal Hall response \cite{alex5a,alex6,alex1}. In some respects, it is analogous to quantum anomalous Hall effect in electronic systems, but requires a temperature gradient and a heat current. From linear response theory,  the general formula for thermal Hall conductivity of spin excitations $\kappa_{xy}$  can be derived, see Ref.~\cite{shin1}.  The low-temperature dependence of  $\kappa_{xy}$ for the present model is plotted in Fig. \ref{N_th}(a) with the parameter values of KFe$_3$(OH)$_{6}$(SO$_{4}$)$_2$ \cite{sup1a}.  At zero magnetic field there is no thermal Hall effect  in accordance with the analysis of topological spin waves and edge modes discussed above.  The crucial feature of this model is that for zero DMI $\boldsymbol{\mathcal{D}}_{ij}=0$, the NNN coupling $\mathcal J_2> 0$ also stabilizes the $\bold q=0$ coplanar structure  as mentioned above \cite{har1} and the thermal Hall effect is present as shown in Fig. \ref{N_th}(b). In fact, an easy-plane anisotropy also has a SOC origin. In the XXZ kagom\'e antiferromagnet  it selects the  $\bold q=0$ LRO without the DMI \cite{cher}.  In this case we also find that topological spin waves and chiral edge modes still persist (see Ref.~\cite{SM}).  These results have established that a TRS-broken chiral spin texture can lead to  nontrivial topological spin excitations in frustrated magnets as opposed to previous studies in which the DMI is the primary source of topological spin excitations \cite{shin1, alex1, alex1a, alex0, alex2,alex5,alex4,bol, sol1,sol, alex5a, alex6,  sol4,  mok,su,fyl,lifa,jud}.

\textit{Conclusion}--.
We have shown that  geometrical spin frustration arising from kagom\'e antiferromagnets can lead to field-induced scalar spin chirality even in the absence of DMI.  The field-induced scalar chirality provides topological spin excitations and thermal Hall response applicable to different frustrated kagom\'e magnets.  These features can be probed by inelastic neutron scattering experiments. In particular, frustrated kagom\'e jarosites such as  KCr$_3$(OH)$_6$(SO4)$_2$ \cite{gro2,gro4} and KFe$_3$(OH)$_{6}$(SO$_{4}$)$_2$ \cite{sup1a}  meet all the requirements predicted in this report. The presence of scalar spin chirality suggests that there is a possibility that  the thermal Hall effect will be present in frustrated kagom\'e QSL materials such as herbertsmithite ZnCu$_3$(OH)$_6$Cl$_2$. Although this compound has a sizeable DMI ($\mathcal D_z/\mathcal J =0.08$) \cite{zor}, the ground state is considered as a QSL with spinon continuum excitations \cite{tia}. However, the QSL phase in herbertsmithite is fragile in the presence of a magnetic field of about 2 T \cite{jeo} and  a pressure of 2.5 GPa \cite{koz}. The chromium compound Ca$_{10}$Cr$_7$O$_{28}$ has also been shown as a QSL candidate \cite{balz}, but also develops a magnetic order in the presence of a magnetic field \cite{balz}.  

In the disordered QSL regime scalar spin chirality can be spontaneously developed as a macroscopic order parameter with broken TRS even without an applied magnetic field \cite{kal,wen,bas,bau}, whereas in the ordered or frozen regime a magnetic field or pressure can equally induce scalar spin chirality as shown in this report. Therefore, we believe that a finite thermal Hall effect in these magnets should be attributed to scalar spin chirality as opposed to the DMI.  A similar effect  in frustrated electronic (metallic) magnets   is known as topological or spontaneous Hall effect \cite{ele0,ele1,ele2,ele, ele3} with or without the magnetic field respectively. The present model is an analog of this effect in quantum magnets with charge-neutral magnetic spin excitations. The main result of  this report is that a combination of geometrical spin frustration and kagom\'e antiferromagnets can realize thermal Hall effect without the need of DMI or SOC.  Furthermore, it would be interesting to probe the analogs of ``Dirac semimetal'' in quantum magnets as pointed out here. The chiral edge modes have not been measured at the moment and they require edge sensitive methods such as  light \cite{luuk} or electronic \cite{kha} scattering method. We recently became aware of a recent experimental report of thermal Hall response in frustrated distorted kagom\'e volborthite  at   $15~\text{T}$ with no signal of DMI \cite{wat}. The mechanism that gives rise to thermal Hall response in this material was not explicitly identified.  In addition, the exact model Hamiltonian for  volborthite and its parameter values are very controversial.  The results presented here apply specifically to undistorted kagom\'e antiferromagnets. In this regard, the present study will be important  in upcoming experimental studies of finite thermal Hall response in frustrated kagom\'e  magnets.

\textit{Acknowledgements}--.
   Research at Perimeter Institute is supported by the Government of Canada through Industry Canada and by the Province of Ontario through the Ministry of Research
and Innovation.


\begin{thebibliography}{99}
\bibitem{alex0}
H. Katsura, N. Nagaosa, and P. A. Lee,   \prl  {\bf 104},  066403 (2010).

 \bibitem{alex1}
Y. Onose, T. Ideue, H. Katsura, Y. Shiomi, N. Nagaosa, Y. Tokura,  Science  { \bf 329}, 297 (2010).
 \bibitem{alex2}
R. Matsumoto and S. Murakami, \prl {\bf 106}, 197202 (2011). {\it ibid} \prb {\bf 84}, 184406 (2011).
  \bibitem{alex1a}
T. Ideue, Y. Onose, H. Katsura, Y. Shiomi, S. Ishiwata, N. Nagaosa, and Y. Tokura, Phys. Rev. B. {\bf 85}, 134411 (2012).
 \bibitem{lifa}
L. Zhang, J. Ren, J. -S. Wang, and B. Li,  Phys. Rev. B {\bf 87}, 144101 (2013).
 \bibitem{alex4}
 A. Mook, J. Henk and I. Mertig , \prb {\bf 90}, 024412 (2014).{\it ibid}  \prb {\bf 89}, 134409 (2014).
   \bibitem{shin1}
R. Matsumoto,  R. Shindou,  and  S. Murakami, Phys. Rev. B 89, 054420 (2014).



 \bibitem{alex6} 
 M. Hirschberger, R. Chisnell, Y. S. Lee, and N. P.  Ong,  \prl {\bf 115}, 106603 (2015).
  \bibitem{alex5a}
R. Chisnell, J. S.  Helton, D. E.  Freedman, D. K.  Singh, R. I.  Bewley, D. G.  Nocera, and Y. S.  Lee,  \prl {\bf 115}, 147201  (2015).
 \bibitem{alex5}
H. Lee, J. H. Han, and  P. A. Lee,    \prb  {\bf 91},  125413 (2015).
  \bibitem{bol}
D. Boldrin, B. Fak, M. Enderle, S. Bieri, J. Ollivier, S. Rols, P. Manuel, and A. S. Wills,  Phys. Rev. B 91, 220408(R) (2015).
\bibitem{sol}
  S. A. Owerre,  J. Phys.: Condens. Matter 28, 386001 (2016).
 \bibitem{sol1}
  S. A. Owerre, J. Appl. Phys. {\bf 120}, 043903 (2016).
\bibitem{fyl}
F. -Y. Li, Y. -D. Li, Y. B. Kim, L. Balents, Y. Yu and G.  Chen, Nat. Commun. {\bf 7}, 12691 (2016).
 \bibitem{mok}
  A. Mook, J. Henk,  and I. Mertig,   Phys. Rev. Lett. {\bf 117}, 157204 (2016).
 \bibitem{su}
 S. Ying, X. S. Wang,  and  X. R.  Wang,   arXiv:1609.01500 (2016).
  \bibitem{sol4}
 S. A. Owerre,  arXiv:1608.00545 (2016).
 
\bibitem{pho}
  C. Kane, and  T. C.  Lubensky,  Nat. Phys. {\bf 10}, 39 (2014).
\bibitem{pho1}
 B. G. Chen,  N. Upadhyaya , and   V. Vitelli,  Proc. Natl. Acad.
Sci. U.S.A. {\bf 111}, 13004 (2014).
\bibitem{pho2}
 J. Paulose, B. G. Chen, and  V. Vitelli,   Nat. Phys. {\bf 11}, 153 (2015).
\bibitem{pho3}
J. Paulose,  A. S. Meeussen and  V. Vitelli,  Proc. Natl. Acad. Sci. U.S.A. {\bf 112}, 7639 (2015).
\bibitem{pho4}
D. Z. Rocklin, B. Gin-ge Chen, M. Falk, V. Vitelli, and T. C. Lubensky,  Phys. Rev. Lett. {\bf 116}, 135503 (2016).
\bibitem{pho5}

O. Stenull, C. L.  Kane, and T. C.  Lubensky, Phys. Rev. Lett. {\bf 117}, 068001 (2016).
\bibitem{dm}
 I. Dzyaloshinsky,     J. Phys. Chem. Solids {\bf 4}, 241 (1958).
 \bibitem{dm2}
T. Moriya,    Phys. Rev. {\bf 120}, 91 (1960).

\bibitem{Sav}
L. Balents,   Nature {\bf 464}, 199 (2010).
 \bibitem{nor}
M. R. Norman,  Rev. Mod. Phys. {\bf 88}, 041002 (2016).
 
\bibitem{ran}
Y. Ran,  {\it et al}.   Phys. Rev. Lett. 98, 117205 (2007).

 \bibitem{men1}
 M. Elhajal, B. Canals,  and C. Lacroix,   Phys. Rev. B {\bf 66},
014422 (2002).
 \bibitem{men3}
O. C\'epas, C. M. Fong, P. W. Leung, and C. Lhuillier,  Phys. Rev. B {\bf 78}, 140405(R)  (2008).
\bibitem{sup1a}
D. Grohol, K. Matan, J. H. Cho, S.-H. Lee, J. W. Lynn, D. G. Nocera, Y. S. Lee,  Nature Materials {\bf 4}, 323 (2005). 
\bibitem{men4a}
K. Matan, D. Grohol, D. G. Nocera, T. Yildirim, A. B. Harris, S. H. Lee, S. E. Nagler, and Y. S. Lee,  Phys. Rev. Lett. {\bf 96}, 247201 (2006).

\bibitem{gro2}
S.-H. Lee, C. Broholm, M. F. Collins, L. Heller, A. P. Ramirez, Ch. Kloc, E. Bucher, R. W. Erwin, and N. Lacevic,  Phys. Rev. B {\bf 56}, 8091 (1997).
\bibitem{gro4}
T. Inami, T. Morimoto, M. Nishiyama, S. Maegawa, Y. Oka, and H. Okumura,  Phys. Rev. B {\bf 64}, 054421 (2001).
 \bibitem{zor}
A. Zorko, S. Nellutla, J. van Tol, L. C. Brunel, F. Bert, F. Duc, J.-C. Trombe, M. A. de Vries, A. Harrison, and P. Mendels,  Phys. Rev. Lett. {\bf 101}, 026405 (2008).
 \bibitem{balz}
C. Balz, B. Lake, J. Reuther, H. Luetkens, R. Sch\"{o}nemann, Th. Herrmannsd\"{o}rfer, Y. Singh, A.T.M. Nazmul Islam, E. M. Wheeler, J. A. Rodriguez-Rivera, T. Guidi, G. G. Simeoni, C. Baines, and H. Ryll,  Nature Phys. {\bf 12}, 942 (2016).
 \bibitem{jeo}
M. Jeong, F. Bert, P. Mendels, F. Duc, J. C. Trombe, M. A. de Vries, and A. Harrison, Phys. Rev. Lett. {\bf 107}, 237201 (2011).
   \bibitem{koz}
D. P. Kozlenko, A. F. Kusmartseva, E. V. Lukin, D. A. Keen, W. G. Marshall, M. A. de Vries, and K. V. Kamenev, Phys. Rev. Lett. {\bf 108}, 187207 (2012).

  \bibitem{kal}
V. Kalmeyer and  R. B.  Laughlin,  Phys. Rev. Lett. {\bf 59}, 2095 (1987).
 \bibitem{wen}
  X. G. Wen,    F. Wilczek,  and A. Zee,    Phys. Rev. B {\bf 39}, 11413 (1989).
  \bibitem{bas}
  G.  Baskaran,  Phys. Rev. Lett. {\bf 63}, 2524 (1989).
  \bibitem{bau}
B. Bauer, L. Cincio, B. P. Keller, M. Dolfi, G. Vidal, S. Trebst, and  A. W. W. Ludwig, Nature Communications {\bf 5}, 5137 (2014).
  \bibitem{magn}
  A. V. Chumak,  V. I. Vasyuchka, Serga,  A. A. Hillebrands,   Nature Phys. {\bf 11}, 453 (2015).
  \bibitem{jud}
  J. Romh\'anyi,  K. Penc, and    R. Ganesh,  Nature Communications {\bf 6}, 6805 (2015).
  \bibitem{SM}
 See supplemental material, arXiv:1609.03563.  
\bibitem{har1}
 A. B. Harris, C.  Kallin, and   A. J. Berlinsky,  Phys. Rev. B 45, 2899 (1992).

 \bibitem{owee}
 S. A. Owerre,  A. A. Burkov , and R. G. Melko,   \prb  {\bf 93}, 144402 (2016). 

\bibitem{ste}
M. Y. Steve, and L. K. Charles   \prl  {\bf 115}, 126803 (2015).
\bibitem{cher}
 A. L. Chernyshev  and M. E. Zhitomirsky,   \prl {\bf 113}, 237202 (2014).
 \bibitem{gong}
S. -S. Gong, W. Zhu, L. Balents, and D. N. Sheng,  Phys. Rev. B 91, 075112 (2015).
\bibitem{hp}
T. Holstein  and H. Primakoff,    Phys. Rev. {\bf 58}, 1098  (1940).
 \bibitem{tia}
T. -H. Han, J. S. Helton, S. Chu, D. G. Nocera, J. A. Rodriguez-Rivera, C. Broholm, Y. S. Lee,  Nature {\bf 492}, 406 (2012).
 \bibitem{ele1} 
Y. Taguchi, Y. Oohara, H. Yoshizawa, N. Nagaosa, Y. Tokura,  Science {\bf 291}, 2573 (2001). 
\bibitem{ele0}
Y. Machida, S. Nakatsuji, Y. Maeno, T. Tayama, T. Sakakibara, and S. Onoda,  Phys. Rev. Lett. {\bf 98}, 057203 (2007).
\bibitem{ele}
Y. Machida, S. Nakatsuji, S. Onoda, T. Tayama, and  T. Sakakibara,  Nature, {\bf 463}, 210 (2008).
 \bibitem{ele3}
C. S\"{u}rgers, G. Fischer, P. Winkel, and  H. v. L\"{o}hneysen,  Nature Commun. {\bf 5}, 3400 (2014).
 \bibitem{ele2} 
J. Zhou, Q. -F.  Liang, H.  Weng, Y. B. Chen, S. -H.  Yao, Y. -F.  Chen, J.  Dong, and G. -Yu Guo, Phys. Rev. Lett. 116, 256601 (2016).


 \bibitem{luuk}
L. J. P. Ament, M. v. Veenendaal, Th. P. Devereaux, J. P. Hill, and J. van den Brink, Rev. Mod. Phys. {\bf 83}, 705 (2011).
\bibitem{kha}
K. Zakeri,  Physics Reports {\bf 545}, 47 (2014).
\bibitem{wat}
D. Watanabe,  K. Sugiia, M. Shimozawaa, Y. Suzukia, T. Yajimaa, H. Ishikawaa, Z.  Hiroia, T.  Shibauchic, Y.  Matsudab, and M.  Yamashita,  Proc. Natl. Acad. Sci. USA {\bf 113}, 8653 (2016).


 \end{thebibliography}
\end{document}